%% file: main.tex
\documentclass[11pt]{article}
\usepackage{pepadun}
\usepackage{lipsum} 
\usepackage{graphicx}
\usepackage{xcolor}
\usepackage{subcaption}
\usepackage{hhline}
\usepackage{amsfonts}

\title{Continual Learning via Ensemble-Based Depth-Wise Masked Autoencoders for Data Quality Monitoring in High-Energy Physics}
\author{%
    \begin{tabular}{c}
        \textsuperscript{1}Dale Julson, 
        \textsuperscript{2}Eric Reinhardt, 
        \textsuperscript{3}Andrii Krutsylo,
        \textsuperscript{2}Resham Sohal,
        \textsuperscript{2}Guillermo Fidalgo,
        \textsuperscript{2}Sergei Gleyzer,
        \textsuperscript{2}Emanuele Usai, \\
        \textbf{The CMS HCAL Collaboration}, 
        \\[6pt]
        \textsuperscript{1}Cerium Laboratories, Austin, TX 78741, USA \\
        \textsuperscript{2}Department of Physics, University of Alabama, Tuscaloosa, AL 35487, USA \\
        \textsuperscript{3}Institute of Computer Science, Polish Academy of Sciences, Warsaw, 02-106, Poland \\ 
        Email: \textsuperscript{1} Dale.Julson@CeriumLabs.com, \\
    \end{tabular}%
}
\date{}

\begin{document}

\maketitle

\begin{abstract}
Machine learning (ML) techniques have been demonstrated to improve the accuracy and efficiency of anomaly detection (AD) when compared to conventional methods. This has led to the adoption of ML for data quality monitoring (DQM) use cases in order to monitor the operation of certain systems to ensure that they are free of undesirable or potentially deleterious anomalies. For applications in the field of High-Energy physics (HEP), where detectors must operate in long-running, harsh environments, ML models used in DQM that have been trained on static datasets are bound to experience degraded performance due to distributional shifts that naturally occur in the incoming data streams, unless directly mitigated via the inclusion of continual ML techniques. This work introduces DepthViT, a lightweight masked autoencoder architecture that employs unique depth-wise embeddings and depth-wise attention, to perform computationally efficient AD tasks. A continual learning framework is developed in which DepthViT models trained on the most recent data streams are ensembled with older models to create a robust overall system which is more resilient to shifts in incoming data streams. When evaluated on occupancy maps from the Compact Muon Solenoid (CMS) hadron calorimeter across multiple data-taking campaigns, the proposed method maintains precision above 98.8\% and a stable ratio of correct anomaly predictions to number of anomalies both under small and large distributional shifts. Beyond HEP, the same ensembling-based continual adaptation strategy can be directly applied to industrial monitoring environments where data also naturally evolve over time. This work therefore presents a path toward adaptive anomaly detection systems capable of sustained operation in dynamic data environments.
\end{abstract}


\section{INTRODUCTION}

In recent decades, machine learning (ML) based approaches have rapidly become the preeminent methods used in a wide range of scientific and engineering fields~\cite{Wang2023}. One such domain that has been revolutionized by the inclusion of ML is anomaly detection (AD). AD is broadly defined as the identification of data points that deviate significantly from the expected distribution of a dataset, and which can often be indicative of errors, failures, or other atypical system level conditions. AD has long been an important effort in many endeavors ranging from manufacturing quality control and scientific discovery, to cybersecurity and even spam filtering in emails. The recent inclusion of ML has allowed for the automation of many AD tasks that previously relied heavily on human operators, resulting in higher throughput and improved accuracy~\cite{Kumari}. One such instances is in the field of high energy physics (HEP), where experiments such as those conducted at CERN's Large Hadron Collider (LHC) have traditionally relied on human operators to monitor incoming particle interaction data to ensure the detectors are producing high-quality data that is suitable for future physics analyses~\cite{refId0}. This effort is referred to as data quality monitoring (DQM). The introduction of ML-based DQM methods in HEP has led to significant improvements in accuracy, often surpassing human-level performance~\cite{pol_2020,BELIS2024100091, HEPMLPaper1, HEPMLPaper2, HEPMLPaper3}. A critical challenge that has received comparatively less attention, however, is the susceptibility of ML-based DQM methods to shifts in the underlying data distribution. HEP detectors operate under extreme conditions, including high radiation environments, cryogenic temperatures, and strong magnetic fields. These conditions can induce gradual or abrupt degradation of detector components, leading to shifts in the incoming data streams, even for data that is non-anomalous~\cite{Strobbe_2017}. For ML-based DQM systems trained on static datasets, such distributional shifts can cause severe performance degradation, a phenomenon we refer to as \textit{model degradation}. As a result of model degradation, anomaly-free data may be incorrectly classified as anomalous. The need to address such challenges motivates the field of continual machine learning (CML), which seeks to develop architectures that can continuously adapt to new data streams while preserving capabilities learned from previous data streams. In this work, we present a novel ML architecture for use in DQM and AD more generally that achieves competitive performance relative to current state-of-the-art approaches in the HEP domain. Our proposed architecture is significantly more lightweight, requiring approximately $1/100^{th}$ of the total parameters compared to existing architectures. Furthermore, we introduce a straightforward ensemble-based CML strategy that effectively mitigates model degradation, leading to improved performance on evolving data streams compared to single-model baselines.

\section{BACKGROUND}

This section provides background information relevant to the hadronic calorimeter (HCAL) sub-detector, existing anomaly detection and DQM schemes, and the need for CML strategies to mitigate the effects of distributional shifts in the incoming data streams.

\subsection{HCAL DETECTOR}

\begin{figure}[!tbp]
  \centering
  \subfloat[]{\includegraphics[width=7cm]{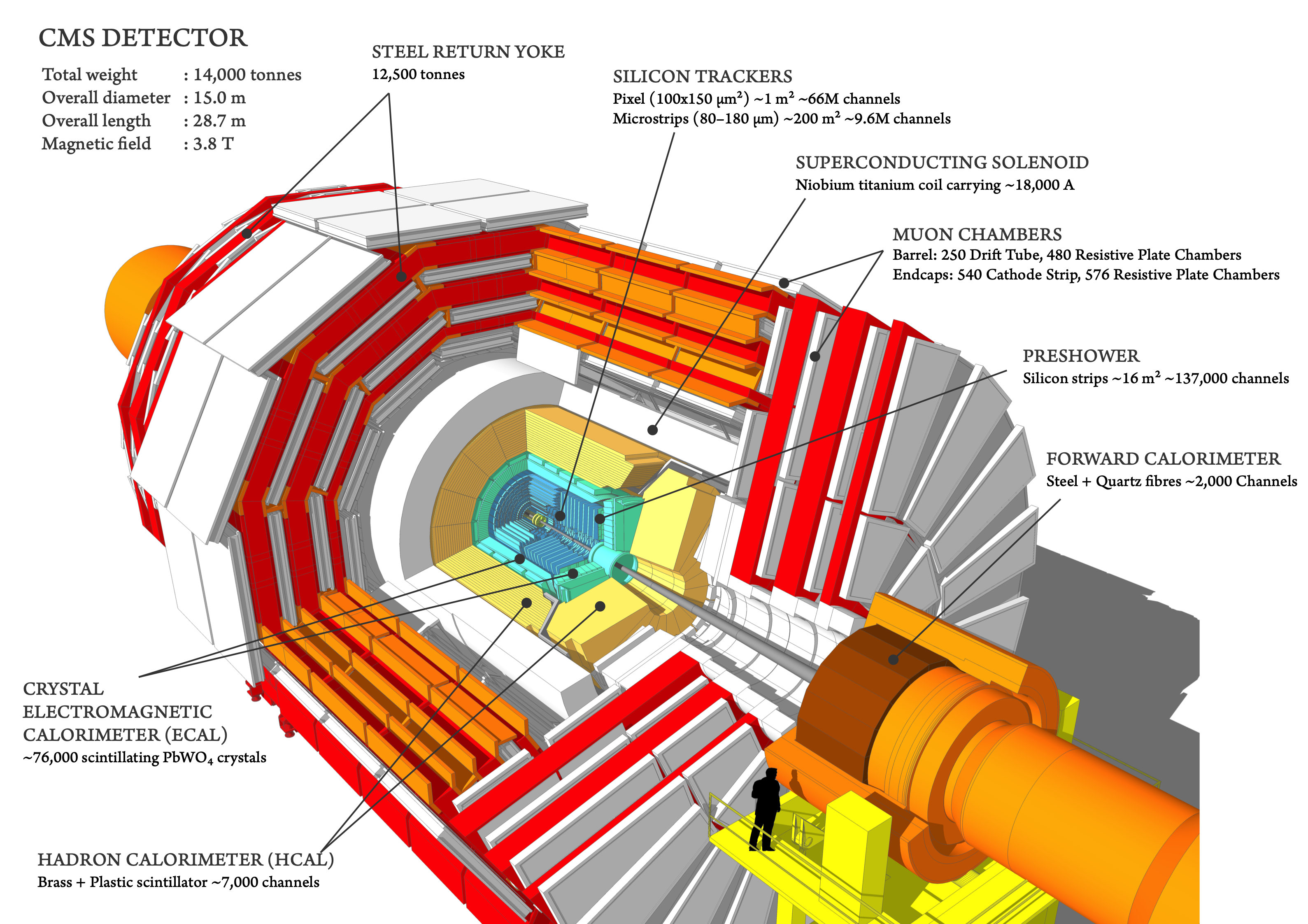}\label{fig:CMS_Detector}}
  \hspace{1em}
  \subfloat[]{\includegraphics[width=8cm]
  {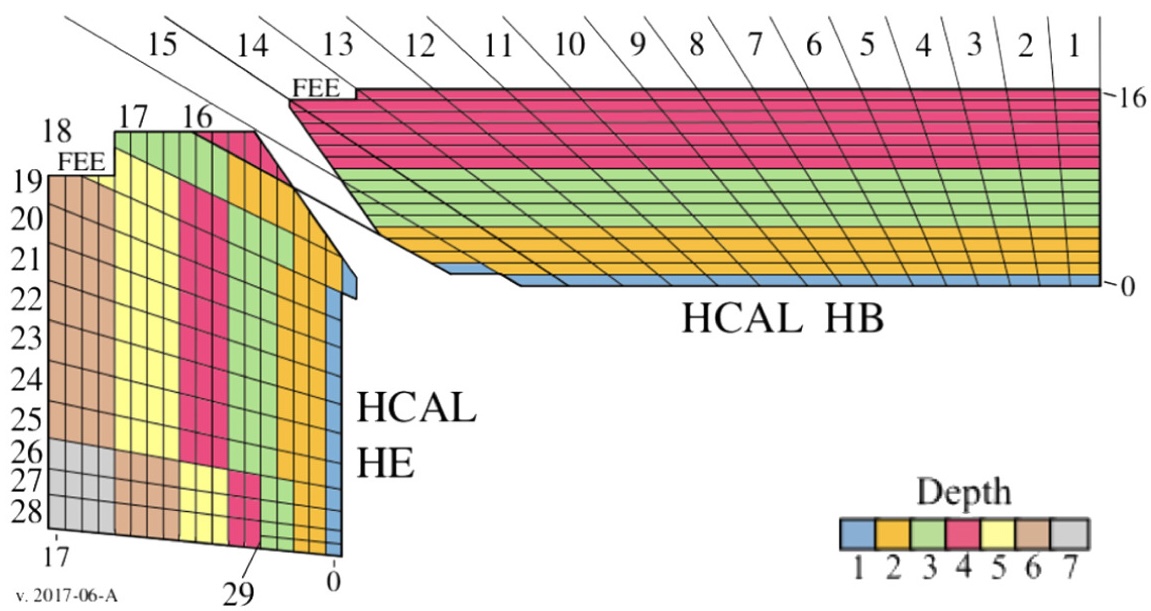}\label{fig:HCAL_Cross_Section}}
  \caption{(a) The CMS detector is comprised of four sub-detectors in addition to a 3.8T superconducting solenoid magnet. Each sub-detector specializes in the detection of different particle species originating from collision interactions. (b) The HCAL sub-detector consists of multiple regions, including the HE and HB regions. The depth and $i\eta$ and $i\phi$ (discretely index pseudorapidity and azimuthal angle) values of these regions are shown here~\cite{CMSDetector}. The HCAL is symmetric in $\phi$. \label{fig:CMS_HCAL}}
\end{figure}

The field of HEP seeks to study the fundamental forces and interactions which underlie the observable universe, often through the use of particle collider experiments. The LHC at CERN is currently the world’s premier facility for collider-based HEP research, hosting several major experiments~\cite{Lyndon}. Among these is the Compact Muon Solenoid (CMS) detector (Fig.~\ref{fig:CMS_Detector}), a general-purpose experiment designed to record the products of proton–proton and heavy-ion collisions for subsequent physics analyses~\cite{The_CMS_Collaboration_2008}. The CMS detector is composed of multiple sub-detectors, including the HCAL (shown in yellow), which measures the energy of charged and neutral hadrons produced in collisions~\cite{Mans:1481837}. The HCAL consists of alternating layers of brass absorber and plastic scintillator, segmented longitudinally into \textit{depths} (Fig.~\ref{fig:HCAL_Cross_Section}). The HCAL is further organized into sections such as the HCAL Endcap (HE) and HCAL Barrel (HB), which are physically distinct portions of the HCAL that can be delineated by their depth, azimuthal angle ($\phi$) and pseudorapidity ($\eta$), which is related to the polar angle $\theta$ by $\eta = -\ln \left [ \tan\left (\frac{\theta}{2}\right)\right]$. During the standard operation of the CMS experiment, where the beam conditions are stable, particle interactions are recorded in periods of approximately 23 seconds, known as \textit{lumisections} (LS). The total number of particle interaction events, which comprise an individual LS, varies between LS due to additional event selection criteria, referred to as \textit{triggers}, which help select for unique and interesting particle interaction events~\cite{Khachatryan_2017}. These LS are then bundled together into larger datasets, referred to as \textit{runs}, which are denoted by a RunID value.

\subsection{DATA QUALITY MONITORING}

The quality of incoming data collected by HEP experiments must be routinely monitored in order to ensure the produced datasets are free of any abnormalities or detector anomalies which may confound future physics analyses~\cite{L_Tuura_2010}. In the case of HCAL, DQM is often performed through the use of DigiOccupancy maps. These maps are three-dimensional histograms that represent the total digitized hits recorded by each individual cell within the HCAL detector during the duration of the LS. The significance of these values is that if an individual cell is not operating properly (i.e. a detector anomaly is present), the DigiOccupancy value will reflect this. Common anomalies that can occur are the presence of $\textit{dead channels}$, in which the cell records no particle interactions at all, $\textit{degraded channels}$, where the cell under-records particle interactions, and $\textit{hot channels}$, in which the cell either over-records particle interactions, or incorrectly records noise as signal. These anomalies can be transient in nature, occurring during a single LS, or persistent in which they occur across multiple LS. Similarly, they can affect individual cells, or entire groupings of cells, depending on the origin of the anomaly. If these anomalies are allowed to persist uninterrupted or otherwise occur undetected, they may either obscure or mimic the very rare signals which are often of interest to particle physicists, or interfere with event triggers which can lead to the collection of uninteresting data.

\subsection{CONTINUAL MACHINE LEARNING}

CML studies the sequential optimization of a model under nonstationary data distributions. Formally, a learner observes a stream of datasets, denoted as $\{\mathcal D_t\}_{t=1}^{T}$. For data collected by the CMS detector, $t$ corresponds to one LS and the input distribution, denoted $p_t(x)$, corresponds to collected data where $x$ is a $\eta$-$\phi$-depth coordinate and the overall distribution may drift with~$t$. The objective is to maintain high performance over time while preventing \textit{catastrophic forgetting}, which refers to a measurable \textit{performance degradation} on earlier data when the model adapts to new distributions~\cite{MCCLOSKEY1989109}.

Classical solutions involve training a single model using regularization, experience replay, or architectural isolation to mitigate forgetting. However, these solutions are often ineffective when distribution shifts are large, interfering, or unpredictable. Regularization methods such as Elastic Weight Consolidation (EWC)~\cite{kirkpatrick2017ewc} and Synaptic Intelligence (SI)~\cite{zenke2017si} assume that task optima lie in a locally smooth region of the loss landscape and degrade when gradients from new data are strongly misaligned with past importance directions. Replay methods like Gradient Episodic Memory (GEM)~\cite{lopezpaz2017gem} and the exemplar-driven incremental classifier iCaRL~\cite{rebuffi2017icarl} require stored exemplars or synthetic samples whose representativeness deteriorates under heavy distribution drift. Architectural isolation approaches such as Progressive Networks~\cite{rusu2016progressive} and PackNet~\cite{mallya2018packnet} mitigate interference but increase memory or structural cost linearly with the number of tasks. As a result, none of these single-model strategies scale well when shifts are continuous, highly non-stationary, or lack explicit task boundaries.

In contrast, ensemble-based continual learning distributes model capacity across multiple experts. Allowing only a subset to adapt decouples plasticity (rapid adaptation) from stability (knowledge retention), enabling models to specialize in different regions of the data distribution. Below, we review the most relevant ensemble-based CML methods and their trade-offs.

CoMA and its Fisher-weighted variant CoFiMA~\cite{marouf2024weighted} address forgetting through parameter averaging. After each task, the newly fine-tuned model is averaged with the previously consolidated model. For instance, if the weights of the newly fine-tuned model are denoted $\omega_t$ and the weights of the previous model are denoted $\omega_{t-1}$, then the weights of the parameter averaged model denoted by $\omega^*_t$ will be given by $\omega^*_t=\lambda\omega_{t}+(1-\lambda)\omega_{t-1}$, where $\lambda$ is a tunable hyperparameter. In CoFiMA, the averaging is weighted by the Fisher information so that parameters important for earlier tasks change less. The resulting averaged model combines plasticity from the current task with proximity to previous optima, retaining prior knowledge without runtime ensemble cost. These methods rely on the assumption of connected loss basins, which refers to joined regions of the parameter loss landscape that are connected by paths of low loss values. If successive optima are separated by high-loss regions, averaging can instead degrade performance.

While CoMA and CoFiMA preserve information through weight averaging, Subspace Ensemble (SE)~\cite{doan2023beyond} achieves similar stability geometrically. SE constrains model parameters to a low-dimensional affine subspace parameterized by convex combinations (all non-zero and summing to one) of a small set of anchor weights. Sampling a new convex coefficient for each mini-batch yields an implicit ensemble with the runtime cost of a single model. That is to say, if there exists a collection of trained models whose weights are denoted by $\omega_i$, where $i$ is an index that references an individual model, then the SE approach creates a new model that contains weights which are a linear combination of the previous \textit{anchor models}. The new model's weights, $\hat{\omega}$, can be expressed as $\hat{\omega}=\sum^{n}_{i=1}\alpha_i \omega_i$, where $\sum^{n}_{i=1}\alpha_i=1$. A subsequent variant introduces connectivity regularization to maintain compatibility between subspaces across tasks. Thus, SE relaxes CoMA's connectivity assumption by explicitly learning compatible regions in parameter space.

Building on the subspace idea, Expandable Subspace Ensemble (EASE)~\cite{zhou2024expandable} targets class-incremental learning on pretrained backbones. EASE freezes the core network and attaches a lightweight adapter for each new task. Each adapter defines a task-specific feature subspace, and inference combines all adapters through subspace reweighting. A semantic prototype complementation module synthesizes old-class prototypes in each new subspace, updating predictions for old classes without replaying old data. Compared with SE, EASE scales to hundreds of classes and maintains low memory overhead by freezing the shared backbone.

Where EASE adds new experts deterministically per task, Differentiable Encoders \& Ensembles (DE\&E)~\cite{wojcik2023naeocl} route updates dynamically. DE\&E employs a frozen feature extractor and multiple classifiers. A differentiable soft-$k$NN gate selects the most relevant classifiers based on feature-space similarity, and only those receive gradient updates. This concentrates plasticity where it is most needed, without assuming known task boundaries. Predictions from all classifiers are combined through a similarity-weighted vote.

In contrast, Selective Expert Ensemble Diversification (SEED)~\cite{rypesc2024seed} assumes explicit task boundaries and assigns exactly one expert to each task. Expert selection is based on similarity to Gaussian class prototypes, and only the chosen expert is fine-tuned. SEED enhances specialization and diversity across experts but depends on fixed prototypes and cannot easily adapt to continuous domain shifts.

When task boundaries are unknown, the Evolved Mixture Model (EEM)~\cite{ye2022evolved} detects novel distributions autonomously. It monitors the Hilbert–Schmidt Independence Criterion between each expert’s latent space and current data, spawning a new VAE–classifier pair when dependence falls below a threshold. This dynamic expansion enables unsupervised adaptation to drifting data streams, at the cost of growing model complexity.

Overall, ensemble-based CML methods mitigate performance degradation by either (i) averaging parameters (CoMA, CoFiMA), (ii) constraining or expanding representational subspaces (SE, EASE), or (iii) routing or selecting experts (DE\&E, SEED, EEM). CoMA and CoFiMA are lightweight but rely on smooth loss connectivity; SE and EASE improve geometric stability with additional parameters; DE\&E and SEED enhance specialization through selective updating yet require gating or task information; and EEM adapts autonomously but expands the ensemble over time. These trade-offs — between runtime cost, memory footprint, reliance on task boundaries, and resilience to unlabelled distribution drift — define the design space for ensemble continual learning and motivate the approach developed in this work.

\section{DATASET DESCRIPTION}

The datasets used in this study consist of DigiOccupancy maps recorded during the 2018 (Run2) and 2022 (Run3) proton-proton collision data-taking campaigns at the CMS experiment. These datasets have been certified as suitable for physics analyses by CMS DQM and are registered as ``\textit{Golden JSON}''~\cite{Golden_JSON}. The 2018 dataset represents a subset of all data recorded by the detector that year, comprising 6 runs collected over a 21-day period (an average of 3.5 days between runs). Similarly, the 2022 dataset is a limited selection of 20 runs spanning a 139-day period (an average of approximately one week between runs). Each DigiOccupancy map is partitioned into 3D spatial coordinates denoted by $i\eta\in[-32,32]$, $i\phi\in[1,72]$, and $depth\in[1,7]$. Here, $i\eta$ is proportional to the pseudorapidity ($\eta$) which represents the angle relative to the beamline, $i\phi$ is proportional to the azimuthal angle ($\phi$), and $depth$ is approximately the segmented radial component within the HCAL. A mask was applied over the datasets so that only the HE portion of the HCAL sub-detector was considered. The HE was chosen given that the endcap regions exhibit a more complex detector geometry than the HB, including more detector depths, equivalent depths located at physically different distances from the interaction point, etc. (see Fig.~\ref{fig:HCAL_Cross_Section}). Therefore, the HE provides both a challenging and representative test case for the CML-based DQM methods proposed here. Including data from the HB would increase the dataset size but is not expected to alter the conclusions of this study.  

\begin{figure}[!tbp]
  \centering
  \includegraphics[width=16cm]{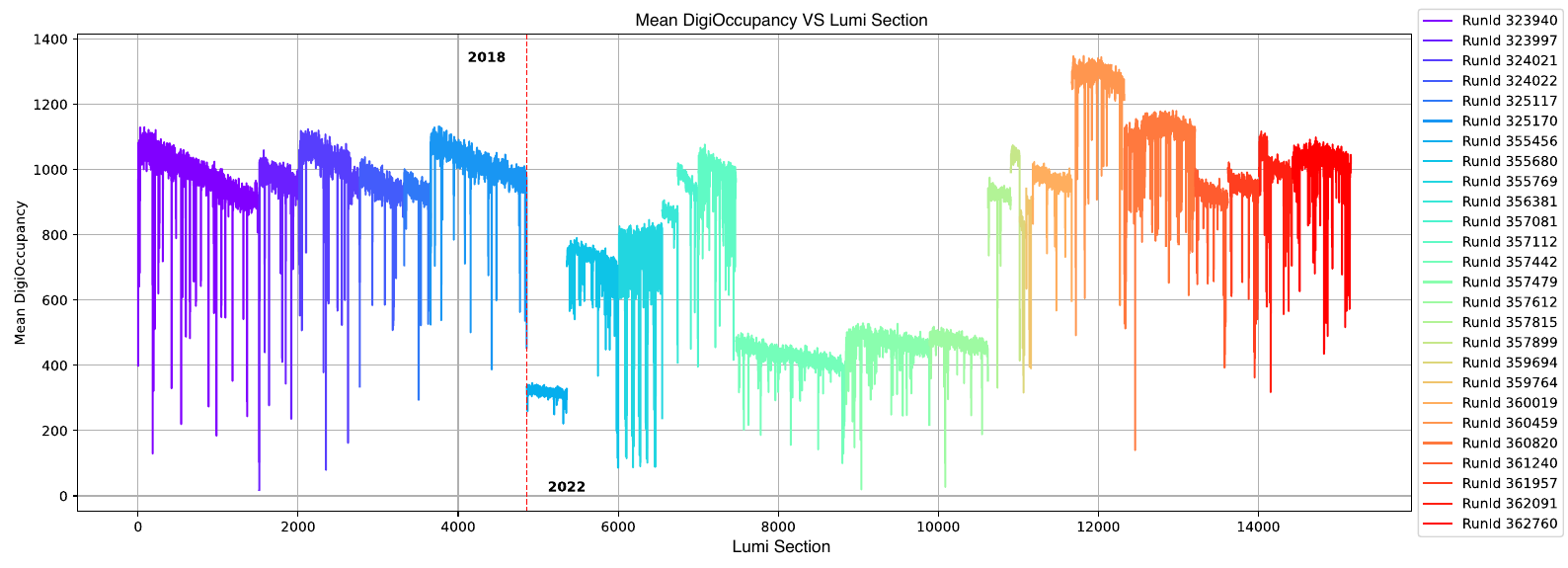}
  \caption{Mean DigiOccupancy versus LS. The red vertical line delineates data originating from 2018 (left) and data originating 2022 (right). The data is temporally ordered both within and between runs, however it should not be assumed that run data is recorded in equally spaced intervals of time.\label{fig:Avg_DigiOccupancy}}
\end{figure}

These datasets exhibit both \textit{small shifts} and \textit{large shifts} in the data distributions. Small shifts can be observed in Fig.~\ref{fig:Avg_DigiOccupancy}, which shows the \textit{mean DigiOccupancy} value across the entire HE on a per-LS basis. The mean DigiOccupancy decreases during the duration of individual runs due to a decrease in the instantaneous luminosity resulting from the many particle interactions that occur~\cite{LHCDesignReport}. Further shifts are observed to occur between runs, as well as between years, caused by changing machine parameters. For instance, there is an approximately 60\% reduction in the mean DigiOccupancy value between the final run considered in the 2018 dataset (RunID 325170), and the first run considered in the 2022 dataset (RunID 355456), and an approximately 188\% increase between the first run considered in the 2022 dataset (RunID 355456) and the final run considered in the 2022 dataset (RunID 362760). As will be further discussed and demonstrated in Sec.~\ref{sec:results}, these small shifts accrue over time and become sufficient to degrade the performance of ML-based DQM models that have been trained on static datasets and which do not employ CML techniques.

\begin{figure}[!tbp]
  \centering
  \subfloat[]{\includegraphics[width=8cm]{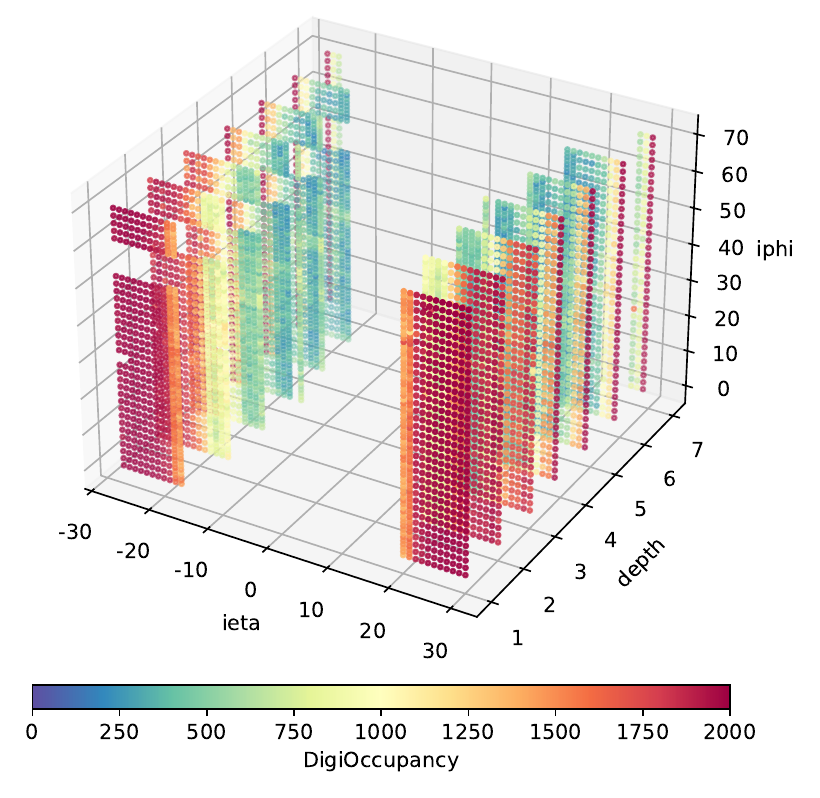}\label{fig:2018_data}}
  \subfloat[]{\includegraphics[width=8cm]{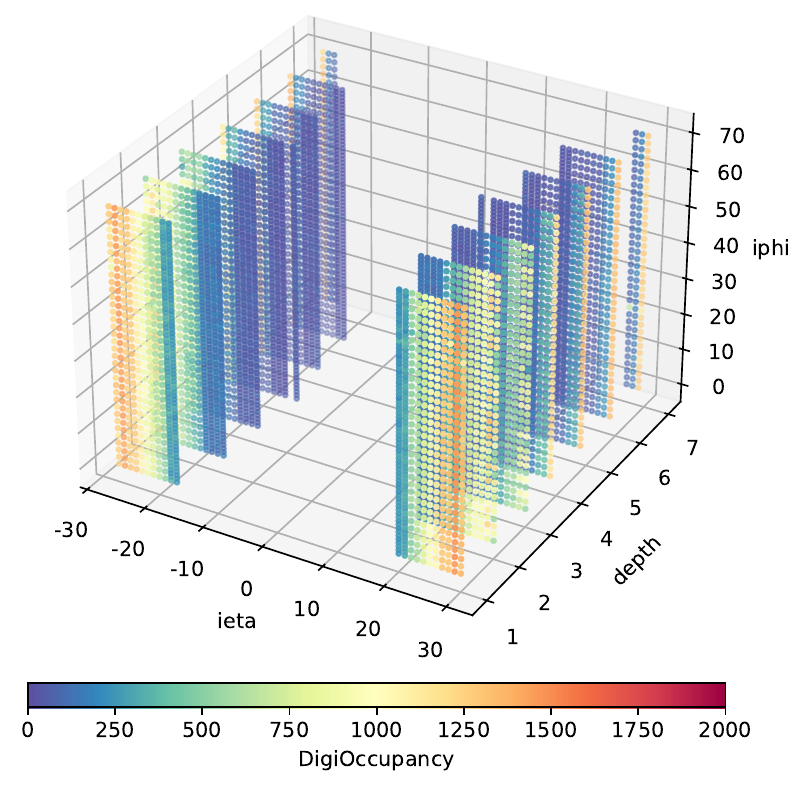}\label{fig:2022_data}}
  \caption{(a) DigiOccupancy map corresponding to \textit{Run325117}, LS 0 (2018). Note the large absence of DigiOccupancy values in the upper left hand corner which resulted from a noted detector failure. (b) DigiOccupancy map corresponding to \textit{Run355456}, LS 0 (2022).\label{fig:datasets}}
\end{figure}

In contrast, Fig.~\ref{fig:datasets} demonstrates an example of a \textit{large shift} that occurred inbetween the 2018 and 2022 data taking campaigns. In particular, during 2018 an approximately 40$^{\circ}$ section in the azimuthal angle ($\phi$) of the HE section of the HCAL lost power, resulting in a large portion of the sub-detector producing no DigiOccupancy values~\cite{HEManomaly}. This is reflected in the upper left portion of the 2018 DigiOccupancy plot shown (left, \textit{Depth} $\in$ [1,7], ~\textit{ieta} $\in$ [-30,-16], ~\textit{iphi} $\in$ [55,62]). As a result, it is expected that a DQM ML model trained on data originating from the 2018 dataset will most likely perform very poorly when evaluated on data originating from the 2022 dataset. Although only a limited number of runs from 2018 and 2022 are used, the presence of these two types of distributional shits in the recorded data presents an excellent basis with which to test and evaluate methods of CML applicable to ML-based DQM. Adding additional runs or additional years of data would primarily increase the amount of nominal data, but is not expected to change the central conclusions regarding model performance and adaptation.

\section{METHODOLOGY\label{sec:methodology}}

\subsection{DepthViT Architecture Overview}

\begin{figure}[!tbp]
  \centering
  \includegraphics[width=16cm]{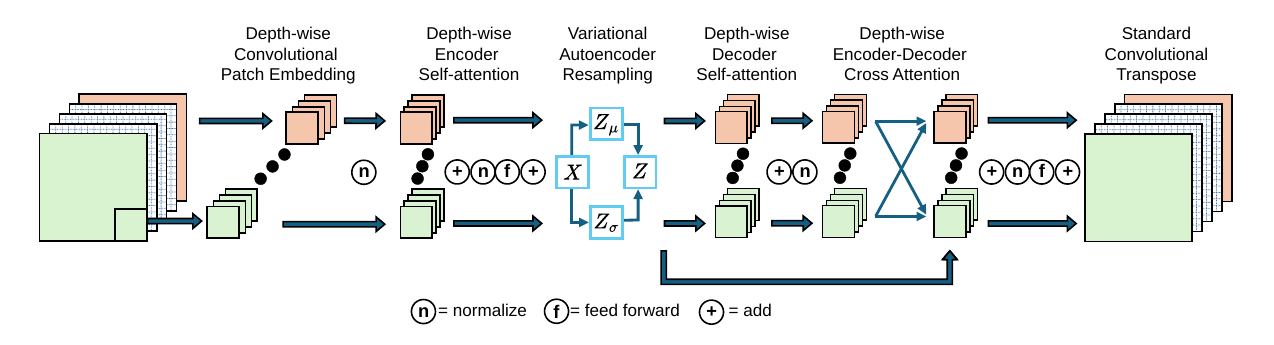}
  \caption{DepthViT variational autoencoder architecture which uses depth-wise convolutional embeddings and depth-wise attention, latent space resampling of the encoder outputs. \label{fig:DepthViT}}
\end{figure}

Here, we introduce a novel masked autoencoder architecture with depth-wise embeddings and depth-wise attention, referred to as \textit{DepthViT}. A graphical representation of the DepthViT architecture is shown in Fig.~\ref{fig:DepthViT}. To better understand the motivation behind this architecture, there is need to further discuss the data distributions produced by CMS DQM systems. Between the individual depths of the HCAL detector, the DigiOccupancy data distributions are not necessarily spatially related. This is because a jet (a collimated spray of hadronic particles) moving through the HCAL detector will initiate further hadronic showers, leading to different shower profiles for different detector depths. The features extracted at one detector depth might not therefore have spatial information that is significant with respect to other detector depths. This is in contrast to a traditional photo image, in which the red, green, and blue channels are expected to all represent the same physical image, leading to contextually rich feature extraction between channels. In this section, the terms \textit{channel} and \textit{depth} are used interchangeably, depending on if the input data is image data or DigiOccupancy maps.

In a typical vision transformer architecture~\cite{visionTransformer}, and in vision models more generally~\cite{maskedautoencoders}, the first step is ordinarily to ``patchify'' the image. This step often involves strided convolutions of the form:
\begin{equation}
    Y_f = \sum_{c=1}^{C} R_{f,c} * X_c,
\end{equation}
where $X \in \mathbb{R}^{H \times W \times C}$ is the input image, $R \in \mathbb{R}^{H'' \times W'' \times C \times F}$ is the convolution kernel (or \textit{filter}), and $Y \in \mathbb{R}^{H' \times W' \times F}$ is the outputted patch. $H$, $W$, $C$, and $F$ refer to height, width, channel number, and filter number, respectively. `$*$' is the convolution operation which is a strided dot product between the overlapping values of $R$ and $X$, which is then summed over the channel dimension. This summation operation immediately mixes information across the input channel dimension. This is appropriate when the images have a channel-wise pixel-level symmetry. This can safely be assumed for RGB color channels in images because the three colors correspond to the same physical point in space, but is not necessarily true in CMS detector data due to different tracks traveling different distances within the detector, and the angled and curved motion of charged particle tracks bending due to the magnetic field. To preserve channel-specific information during patching, the DepthViT architecture uses \textit{depth-wise} convolutional patching, shown in Eq.~\ref{eq:depthconv}. Here the key difference is that the channel dimension is not summed over, hence the $k,c$ dimensionality in place of $f$:

\begin{equation}\label{eq:depthconv}
    Y_{k,c} = R_{k,c} * X_c, \qquad Y \in \mathbb{R}^{H' \times W' \times C \times F}.
\end{equation}

\begin{figure}[!tbp]
  \centering
  \subfloat[]{\includegraphics[width=8cm]{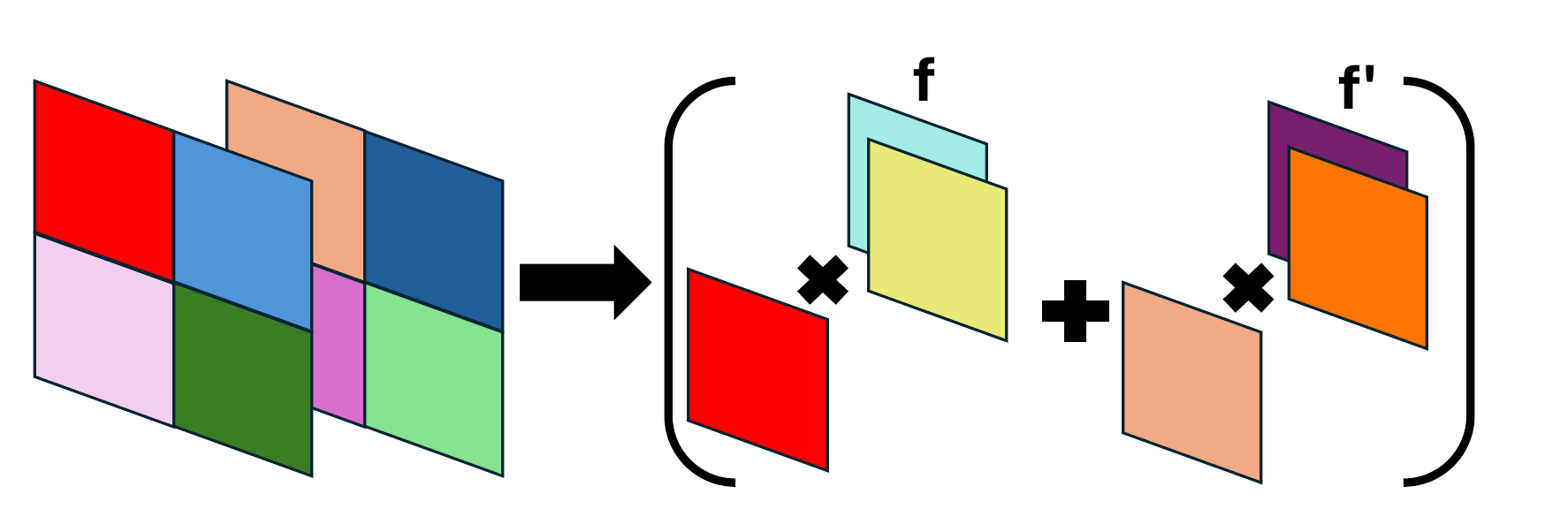}\label{fig:DepthViT_viz1}}
  \subfloat[]{\includegraphics[width=8cm]{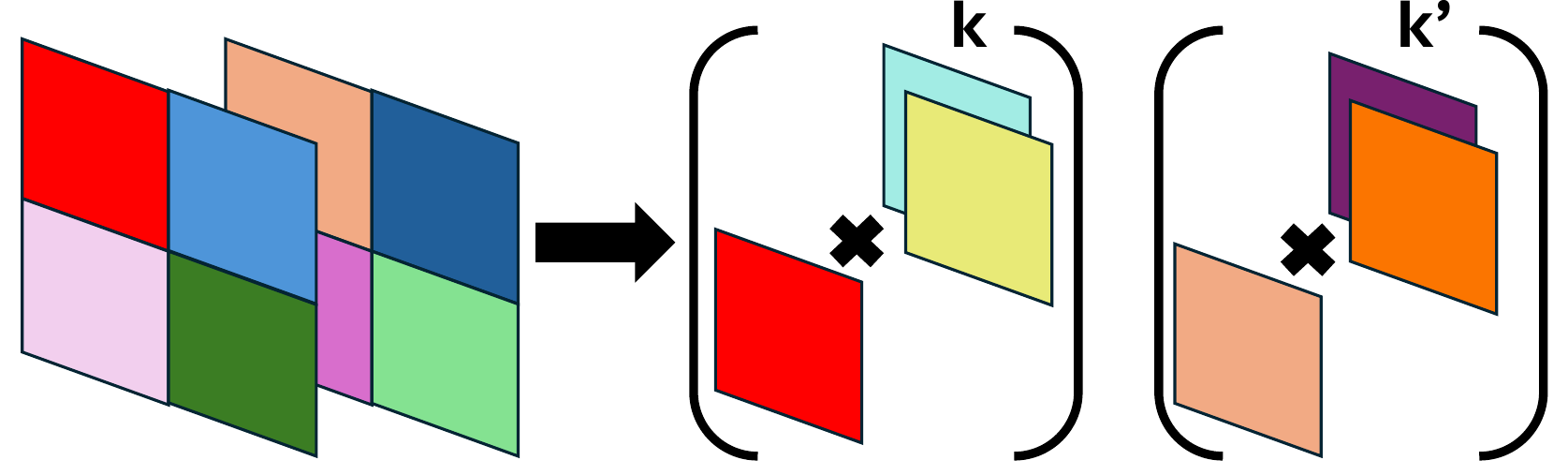}\label{fig:DepthViT_viz2}} \\
  \subfloat[]{\includegraphics[width=8cm]{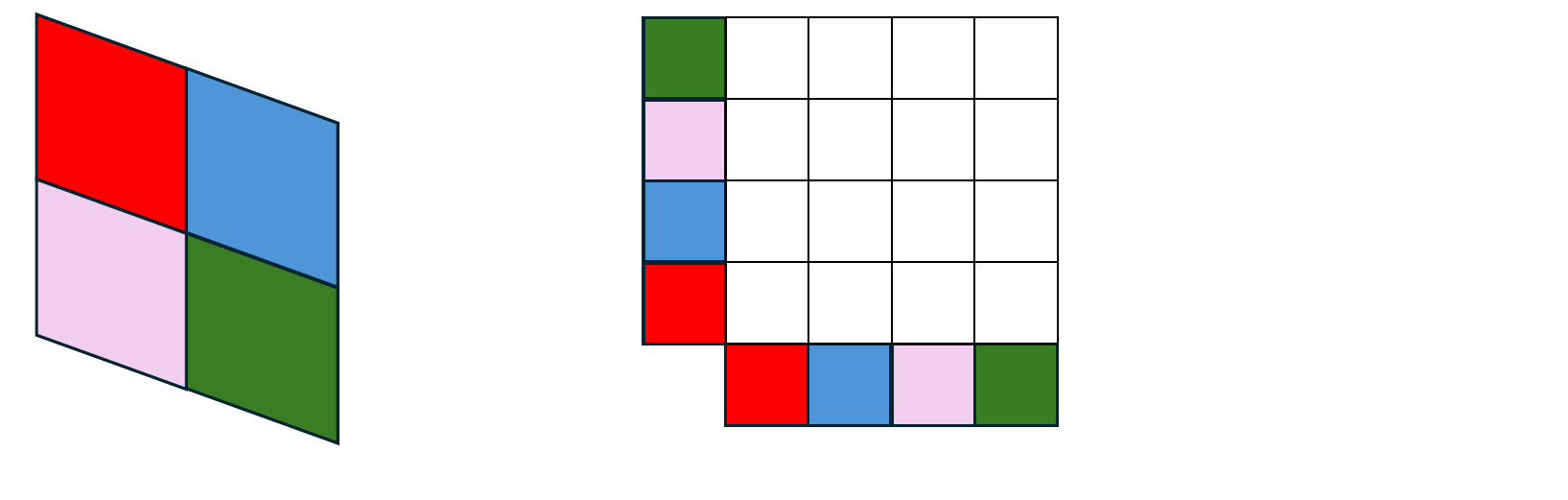}\label{fig:DepthViT_viz3}}
  \subfloat[]{\includegraphics[width=8cm]{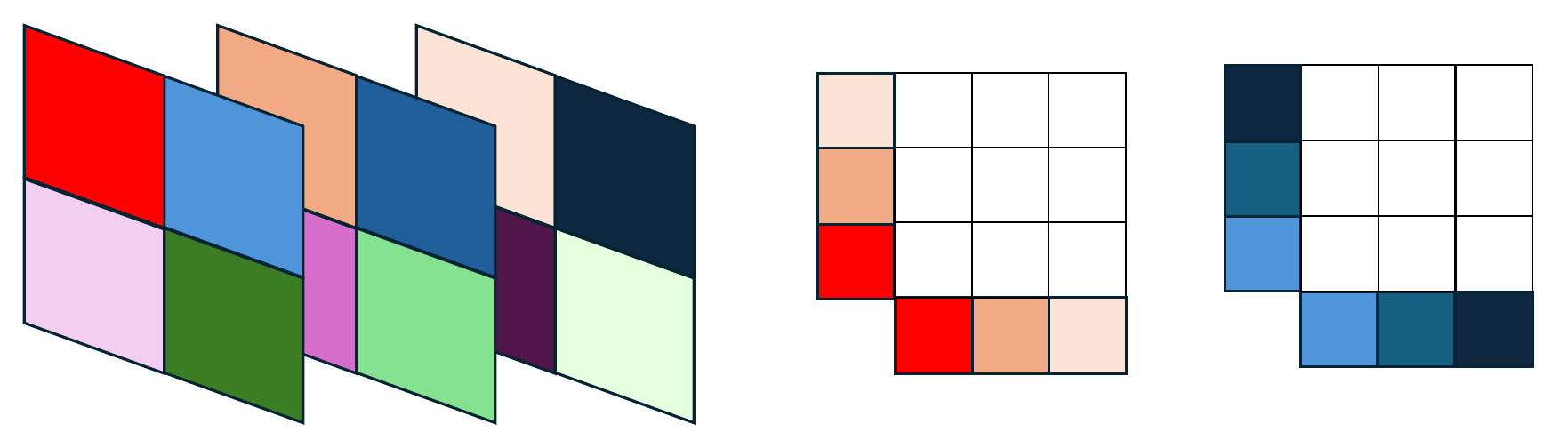}\label{fig:DepthViT_viz4}}
  \caption{(a) Traditional convolution procedure used within vision transformer architectures, where $f$ represents the shared kernel filters. (b) convolution procedure used within the DepthViT architecture, where $k$ and $k'$ represent the filters derived separately for each channel. (c) Traditional attention mechanism used within vision transformer architectures where the attention mechanism attends along the input sequence. (d) Depthwise attention mechanism used within the DepthViT architecture where the attention mechanism instead attends in a depth-wise manner.\label{fig:DepthVit_visualization}}
\end{figure}

This difference is visualized in Fig.~\ref{fig:DepthViT_viz1}, which shows the traditional approach, and Fig.~\ref{fig:DepthViT_viz2}, which shows the approach taken within DepthViT. Because the convolutional kernel weights are shared across all patches from the image, some level of information about channel-specific relationships between patches is retained, while most channel-wise relationships are abstracted away. To reintroduce channel-wise relationships, we additionally introduce a novel, depth-wise attention mechanism. 

A traditional vision transformer attention mechanism involves taking as input a collection of embedded image patches, $Z \in \mathbb{R}^{L \times D}$, where $L$ is the sequence length and $D$ is the embedding dimension. For vision transformers using convolutional patch embeddings, $L$ is the flattened input image sequence length ($L=H'\cdot W'$) and $D$ is equal to the number of kernel filters ($D=F$). Learned linear projection operators $W_Q$, $W_K$, and $W_V$ are then applied along the embedding dimension to produce:
\begin{equation}
    Q = ZW_Q, \qquad K = ZW_K, \qquad V = ZW_V,
\end{equation}
where $Q,K,V \in \mathbb{R}^{L \times D}$ are the query, key, and value projections of the embedded image patches, respectively. Next, an outer product is taken of $Q$ and $K^T$ along the sequence length dimension to compute the scores, $S$:
\begin{equation}
    S_{ij}=\frac{1}{\sqrt{D}}\sum_{r=1}^{D} Q_{ir}K_{jr}, \qquad S \in \mathbb{R}^{L \times L}.
\end{equation}
The scores are then passed through a softmax function to get the attention weights, $A$:
\begin{equation}
    A_{ij}=\frac{\exp(S_{ij})}{\sum_{j'=1}^{L}\exp(S_{ij'})}, \qquad A \in \mathbb{R}^{L \times L}.
\end{equation}
Finally, the attention weights are taken as a dot product along the key and value sequence length dimension to get the attention output which is added back to the original embedded input:
\begin{equation}
    O_{ij} = \sum_{r=1}^{L} A_{ir}V_{rj} \qquad O \in \mathbb{R}^{L \times D}.
\end{equation}
In such an attention framework, the attention weights are considered based on relationships between patches, as visualized in Fig.~\ref{fig:DepthViT_viz3}.
In the DepthViT architecture, the attention mechanism considers relationships between patches along the channel dimension, rather than along the patch dimension. This difference is visually demonstrated in Fig.~\ref{fig:DepthViT_viz4}. Here, because we are using depth-wise convolutional patch embedding, $Z \in \mathbb{R}^{L \times C \times F}$ and $Q,K,V \in \mathbb{R}^{L \times C \times F}$, where $L$ is again the flattened input sequence length ($L=H' \cdot W'$), $C$ is the channel dimension, and $F$ is the number of kernel filters per channel. The scores are then computed by taking an outer product along the channel dimensions:
\begin{equation}
    S_{ijkl} = \frac{1}{\sqrt{F}} Q_{ijl}K_{ikl}, \qquad S \in \mathbb{R}^{L \times C \times C \times F},
\end{equation}
where $i$ indexes over patches, $l$ over filters, and $j$ and $k$ are projections along the channel dimension. The attention maps also have a softmax applied along the channel dimension rather than the sequence dimension.
\begin{equation}
    A_{ijkl} = \frac{\exp(S_{ijkl})} {\sum_{k'=1}^{C}\exp(S_{ijk'l})}, \qquad A \in \mathbb{R}^{L \times C \times C \times F}.
\end{equation}
Finally, the output vectors are computed from an inner product along the key and value channel dimensions with C as the number of channels:
\begin{equation}
    O_{ijl} = \sum_{k=1}^{C} A_{ijkl}V_{ikl}, \qquad O \in \mathbb{R}^{L \times C \times F}.
\end{equation}
This alternative attention mechanism broadcasts along the sequence dimension computing cross-depth attention. This eliminates information sharing between patches in a layer, but can save substantially on parameters for cases where cross-channel symmetries are more significant. Within DepthViT this depth-wise attention is applied in both a self-attention manner, in which $q$, $k$, and $v$ all originate from the same sequence, and in a cross attention manner, in which $q$ originates from the decoded sequence and $k$ and $v$ originate from the sequence before decoding. After the attention reweighting is completed, the images are transformed back to the original dimensions using a standard convolutional transpose which is analogous to an inverse convolution operation which mixes channel information during reconstruction. 

To improve the performance of the model in AD, some subset of all input patches (set to 50\% for this implementation) are randomly masked with a tunable likelihood during training. This masking acts as an information bottleneck to ensure that the model does not learn to simply pass the identity through the transformer encoder and decoder. Similarly, after depth-wise encoding the sequence is passed through an optional light variational bottleneck which encourages smoother, more noise-tolerant latent features. As a result of these novel vision transformer techniques, DepthViT is a lightweight masked autoencoder architecture that relies on approximately 300,000 parameters. In comparison, the smallest default ViT-B/16 \cite{visionTransformer}, commonly used for such tasks, has 86M parameters, and the lightweight Self-Distilled Masked Auto-Encoder \cite{ristea2024selfdistilledmaskedautoencodersefficient} has 3M parameters. In addition, the computational complexity changes from $O(L^2D)$, where $L$ is the number of patches or sequence length and $D$ is the number of filters or embed dimension, to $O(LC^2D)$ which means that these architectures also scale well in compute time when $C\ll L$.

\subsection{DepthViT Based Anomaly Detection}

In this application, AD is performed by computing Z-score values and using them to quantify deviations from nominal detector behavior. This technique is performed in the following manner. A DepthViT model is first trained on pristine detector data which contains no anomalies. The \textit{prediction error} ($\mathrm{pred}_{\mathrm{err}}$), defined as the difference between the model's prediction and the target data, is computed on a per-pixel basis by deploying the model on its associated test data. The mean ($\mu_{\text{err}}$) and standard deviation ($\sigma_{\text{err}}$) of these errors are calculated for each pixel, and used as a baseline performance metric for the model. During inference, a time-window technique is employed, in which $T$ LS are processed together. For each time-window $T$, the Z-score is calculated in the following manner:

\begin{equation}\label{eq:Z_score}
Z=\frac{\text{abs}(\sum_{t=1}^{t=T} \text{pred}_{\text{err}}-T*\mu_{\text{err}})}{\sqrt{T}*\sigma_{\text{err}}},
\end{equation}

where $\text{abs}$ is the absolute value function, and $t$ refers to the individual LS present within time-window $T$ that are summed over. The intuition behind this procedure is that for time-window LS values which do not contain anomalies, it is expected that $\text{pred}_{\text{err}}$ values will represent a sampling from a distribution with mean of $\mu_{\text{err}}$ and therefore the quantity in the numerator of Eq.~\ref{eq:Z_score} will equate to zero. For time-windowed LS which do contain anomalies, $\text{pred}_{\text{err}}$ is no longer sampled from a random distribution but is instead expected to consistently lie either above or below the $\mu_{\text{err}}$ value, implying that over the entire time-windowed period the Z-score will consistently increase in value and improve the capacity to identify anomalies.

\begin{figure}[!tb]
  \centering
  \subfloat[]{\includegraphics[width=8cm]{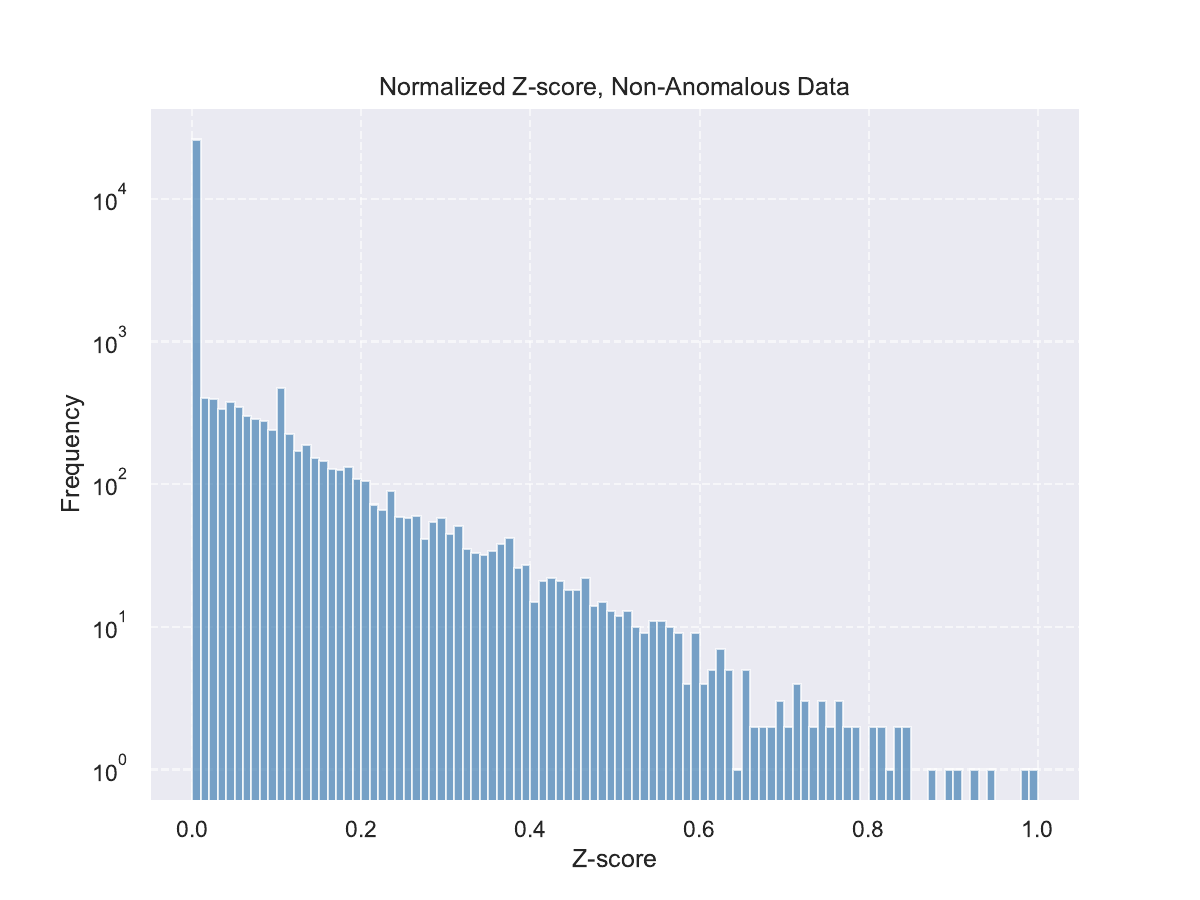}\label{fig:ZScore_no_anomaly}}
  \subfloat[]{\includegraphics[width=8cm]{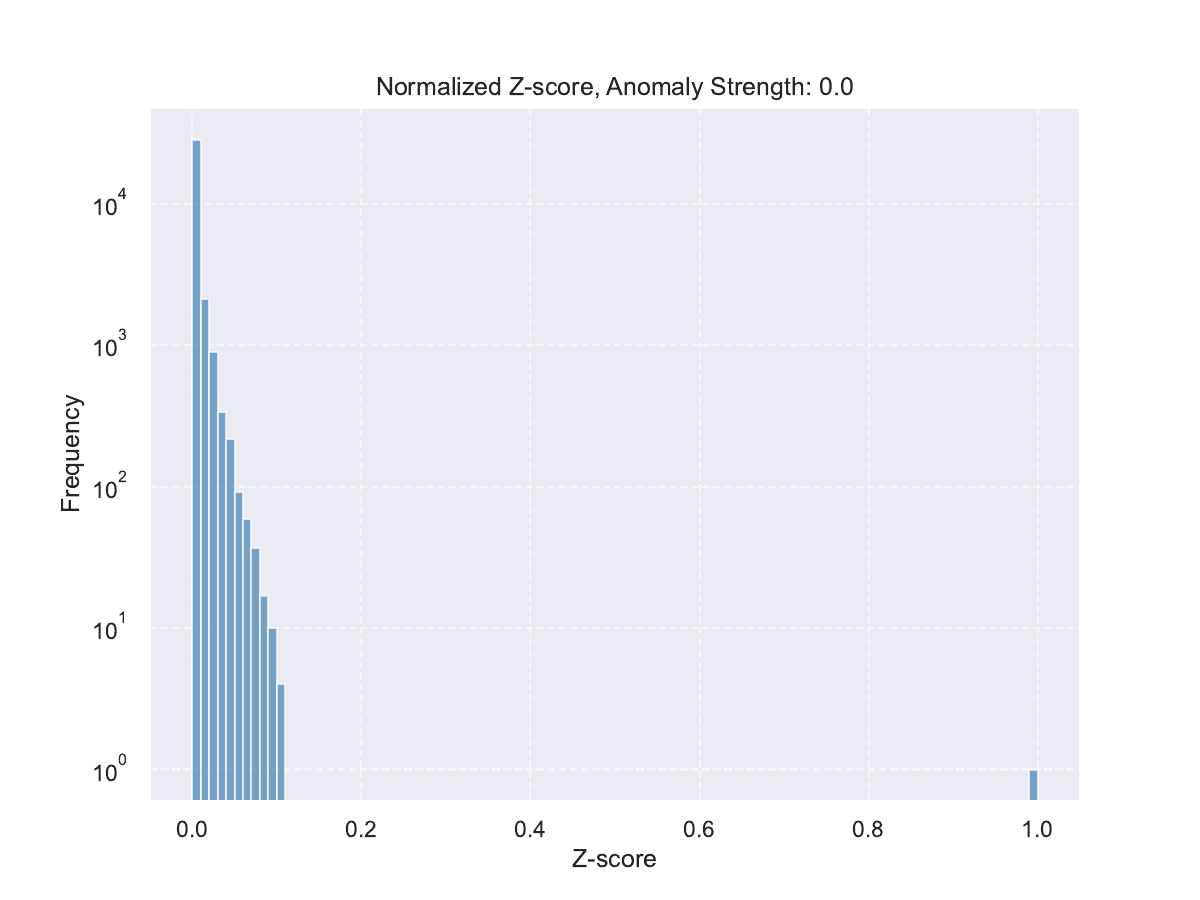}\label{fig:ZScore_with_anomaly}}\\
  \vspace{-1em}
  \subfloat[]{\includegraphics[width=8cm]{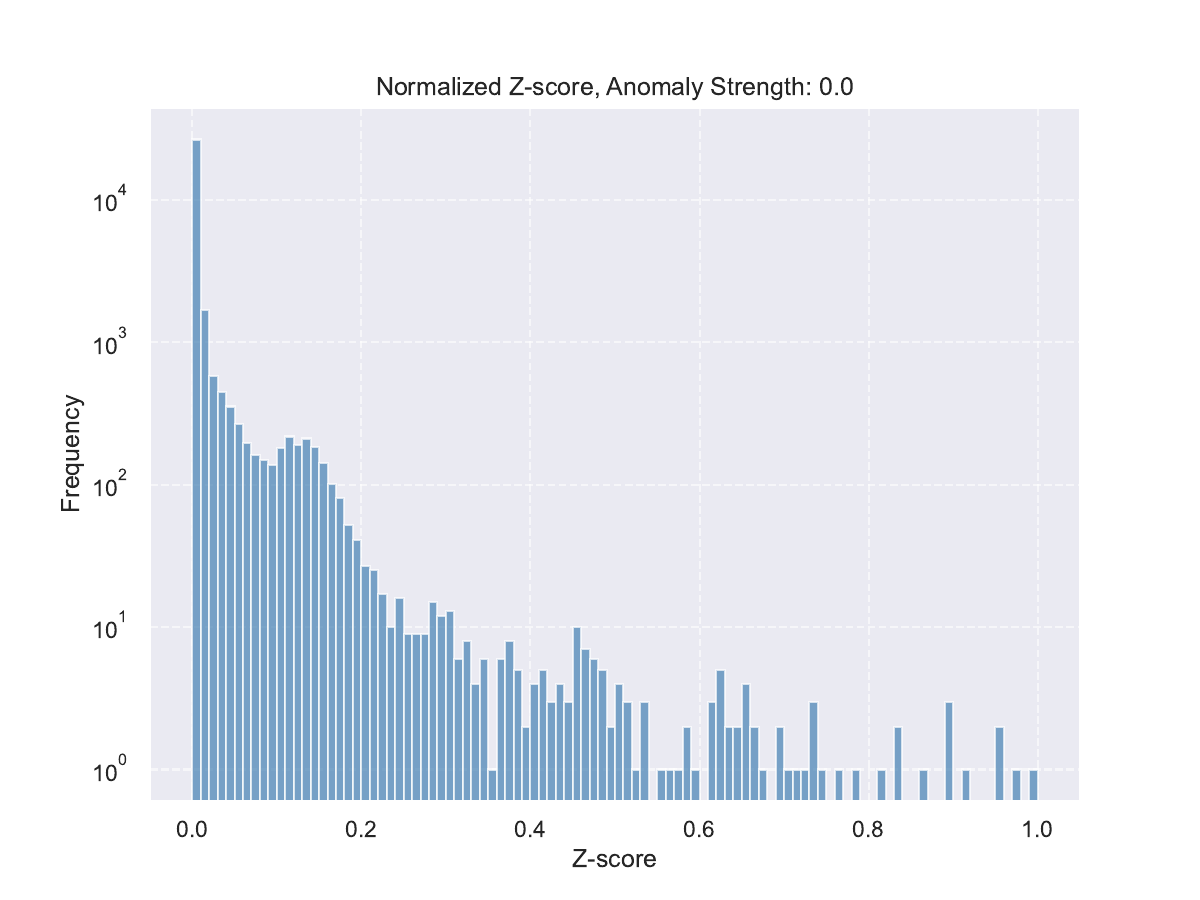}\label{fig:ZScore_with_anomaly_poor_performance}}
  \caption{(a) Normalized Z-score distribution of a well-performing DepthViT model when deployed on non-anomalous data. (b) Normalized Z-score distribution of a well-performing DepthViT model when deployed on anomalous data with anomaly factor of 0.0. (c) Normalized Z-score distribution of a severely degraded DepthViT model when deployed on anomalous data with anomaly factor of 0.0.\label{fig:Z_score_distributions}}
\end{figure}

A variety of different threshold mechanisms were considered for use in determining the presence of anomalies, including a single threshold value, $\theta$, such that if $Z>\theta$ for any Z-value then the data would be labeled anomalous. Ultimately, this single threshold value approach was found to not offer robust performance and instead led to an increased false positive rate (FPR) when deployed on datasets originating from outside the training data. Instead, a \textit{gap-score} method was employed which identified large gaps in the Z-score distribution to indicate the presence of anomalies. For this method, all Z-score values are calculated within a given LS. The Z-score distribution is then normalized to the largest Z-score value, $Z_{1}$, so that the distribution is bounded between $(0,1]$, and the distribution is ordered in ascending order. The gap-score, $G$, is calculated as:
\begin{equation}
    G=Z_{1}-Z_{2},
\end{equation}
where $Z_{2}$ is the second largest Z-score value. For any $G$ value greater than $G_0$ the LS is labeled as anomalous, where $G_0$ is a detection parameter that can be optimized for specific use cases. The use of only the two largest Z-score values in the gap-score calculation is due to only single anomalies per LS being considered in this work. However, the same technique can easily be extrapolated to the presence of multiple simultaneous anomalies by simply calculating $G$ between each sequential pair of ordered Z-score values and identifying any LS with $G>G_0$ as being anomalous. Fig.~\ref{fig:Z_score_distributions} shows the normalized Z-score distribution for a variety of different scenarios including a well-performing model deployed on non-anomalous data (top left), deployed on anomalous data (top right), and a degraded model deployed on anomalous data (bottom). Both the well-performing model deployed on non-anomalous data and the degraded model deployed on anomalous data produce similar distributions, neither of which would be labeled as anomalous. This tendency for a degraded model to label all data as non-anomalous helps reduce the FPR, while a single well-performing model within the overall ensemble of models reduces the False Negative Rate (FNR).

\subsection{Ensembling Technique}

\begin{figure}[!tbp]
  \centering
  \includegraphics[width=16cm]{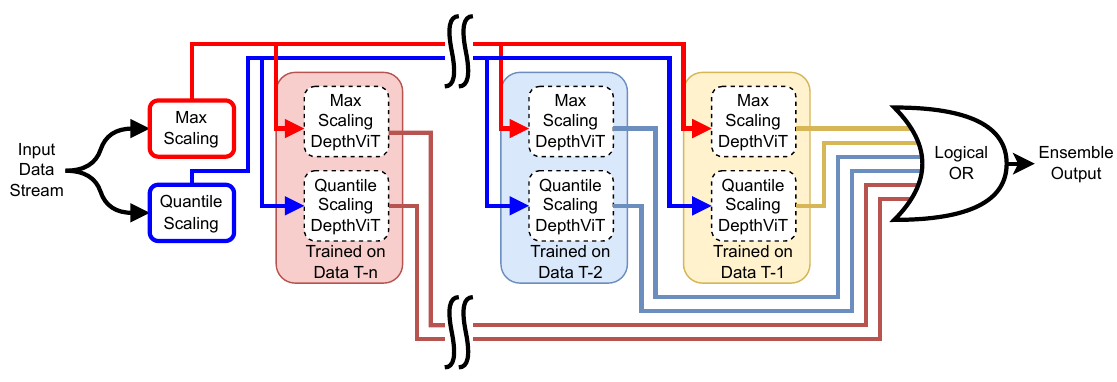}
  \caption{Flow chart detailing the ensembling scheme implemented within this work. Input data is scaled separately using max scaling and quantile scaling operations. The data is then fed in a parallel fashion into individual sub-ensembles which consist of independent DepthViT models trained on max-scaled data and quantile-scaled data respectively. The outputs from all sub-ensembles is combined using a logical OR operation, where if any sub-ensemble labels the data as being anomalous then the entire ensemble output label is anomalous.\label{fig:Flow_chart}}
\end{figure}

The lightweight nature of the DepthViT architecture makes it ideal for use in ensembling methods, which have the potential to constrain memory resources. For this work, a straightforward ensembling strategy was employed and evaluated for its ability to counteract the effects of model degradation. Separate DepthViT models were trained for each run, using training and test datasets derived exclusively from that run. Ensembles of DepthViT models were then deployed in parallel on the same incoming data stream, and per-model Z-score distributions were calculated. These Z-score distributions were fed into the gap-score equation (Eq.~\ref{eq:Z_score}) which produced an anomalous/non-anomalous classification prediction from each model of the overall ensemble. If any individual model classified the data as anomalous, then the ensemble output prediction is an anomalous classification for that LS. This approach is trivially parallelizable, implying it does not necessarily increase the compute-time for systems with sufficient processing power. Anomaly classification tests were performed on test data that were injected with synthetic anomalies. \textit{Anomaly factors} of ``0.0'', ``0.2'', ``0.4'', ``0.6'', ``0.8'', ``1.5'', and ``2.0'' were considered for this study. The location of an anomaly was chosen by randomly selecting with equal probability an individual pixel within the HE, and the DigiOccupancy for that pixel was then multiplied by the anomaly factor value. These anomaly strengths reflect actual detector abnormalities, with 0.0 representing a dead channel, 0.2 to 0.8 representing degraded channels, and 1.5 and 2.0 representing hot channels. Time-windows of varying lengths were considered for both scenarios in which the anomalies were transient or persistent. Two different scaling methods, \textit{quantile} and \textit{max}, were employed in parallel as part of a pre-processing step. This was due to the empirical observation that max scaling provided better sensitivity to anomaly factors $<1.0$, while quantile scaling provided better sensitivity to anomaly factors $>1.0$. Quantile scaling was applied using the following formula:

\begin{equation}\label{eq:quantile_scaling}
Y_{\text{scaled}}=\frac{X-\text{Q}_{25}(X)}{\text{Q}_{75}(X)-\text{Q}_{25}(X)+\epsilon}
\end{equation}

where $X$ represents the DigiOccupancy values for a single LS, $\text{Q}_{25}(X)$ is the sampled 25th percentile value, $\text{Q}_{75}(X)$ is the 75th percentile value, $\epsilon$ is a small value to offset any potential divide-by-zero errors and $Y_{\text{scaled}}$ is the resulting scaled output data. \textit{Max scaling} was applied using:

\begin{equation}\label{eq:max_scaling}
Y_{\text{scaled}}=\frac{X}{X_{max}}
\end{equation}

where again $X$ represents the input DigiOccupancy values for a single LS, and $X_{max}$ is the maximum value within the input array for a single LS. Only one version of the scaled data was fed into an individual DepthViT model, as determined by which scaled data version the model was trained on. A graphical workflow of this process is shown in Fig.~\ref{fig:Flow_chart}. The term \textit{sub-ensemble} will be used here to describe a Max Scaling DepthViT model placed in parallel with a Quantile Scaling DepthViT, and will represent the smallest building block used to create larger ensembled systems.

\section{RESULTS AND DISCUSSION\label{sec:results}}

\subsection{Baseline Model Performance\label{sec:baseline}}

Presented first are baseline results demonstrating the performance of a single DepthViT sub-ensemble model. This model was trained on \textit{Run323940} data which originates from the 2018 dataset. The HE data was first randomized by LS and then divided using a 60\%/20\%/20\% split into \textit{training data}, \textit{test data}, and \textit{validation data}. The $\mu_{\text{err}}$ and $\sigma_{\text{err}}$ values were derived by deploying the trained model on its associated test data. Synthetic anomalies of the previously-described values were injected into the validation data to simulate run-time inference and evaluate performance. Only one anomaly factor was considered at a given time (anomalies of different values were not injected into the same dataset). Optimization studies were performed on a variety of hyperparameters including patch size, number of kernels, masking ratio, and time-window length $T$. The optimal hyperparameter values were selected according to their anomaly detection performance. Varying the patch size did not lead to improved performance, however it was found that decreasing the patch size did lead to longer inference compute times. A gap-score value of $G_0=0.3$ was selected as the optimized threshold for events to be classified as anomalous. Increasing $T$ was found to improve recall performance for persistent anomalies across all anomaly factors, however recall performance was found to degrade as a result for transient anomalies. Precision performance improved for increasing $T$, however, the improvements are marginal except in the shortest transient cases. As a trade-off between these results, a time-window value of $T=5$ was selected as an optimized value. Table~\ref{tab:anomaly_strength_metrics} shows the performance of a DepthViT model using the optimized hyperparameter values across all anomaly factors considered for the metrics of precision, recall, F1 score, FPR, \& FNR. It was observed that the model performed very well, even for subtle anomaly factors (e.g. anomaly factor of 0.8), therefore additional anomaly factors were included in this study.

\begin{equation}
    \textbf{Precision}=\frac{TP}{TP+FP};\textbf{ Recall}=\frac{TP}{TP+FN};\textbf{ F1 Score}=\frac{2}{\frac{1}{\textbf{Recall}}+\frac{1}{\textbf{Precision}}}
\end{equation}
\begin{equation}
    \textbf{FPR}=\frac{FP}{FP+TN};\textbf{ FNR}=\frac{FN}{FN+TP}
\end{equation}

\begin{table}[ht]
\centering
\begin{tabular}{|c|c|c|c|c|c|}
\hline
\textbf{Anomaly Factor} & \textbf{Precision} & \textbf{Recall} & \textbf{F1 Score} & \textbf{FPR} & \textbf{FNR} \\
\hline
0.0 & 1.000 & 1.000 & 1.000 & 0.000 & 0.000 \\ 
0.2 & 1.000 & 1.000 & 1.000 & 0.000 & 0.000 \\ 
0.4 & 1.000 & 1.000 & 1.000 & 0.000 & 0.000 \\
0.6 & 1.000 & 1.000 & 1.000 & 0.000 & 0.000 \\
0.8 & 1.000 & 0.967 & 0.983 & 0.000 & 0.033 \\ 
0.9 & 1.000 & 0.783 & 0.879 & 0.000 & 0.217 \\
1.1 & 1.000 & 0.667 & 0.800 & 0.000 & 0.333 \\
1.25 & 1.000 & 0.983 & 0.992 & 0.000 & 0.017 \\
1.5 & 1.000 & 1.000 & 1.000 & 0.000 & 0.000 \\
2.0 & 1.000 & 1.000 & 1.000 & 0.000 & 0.000 \\
\hline
\end{tabular}
\caption{Relevant performance metrics of a DepthViT model trained on $Run323940$ using optimized hyperparameters (patch size = 12, kernels = 8, mask ratio = 0.5, $T=5$, layers = 5, latent dimension = 56, $G>0.3$). Inference was performed using synthetic anomalies of the listed strengths that were injected into the associated test dataset.}
\label{tab:anomaly_strength_metrics}
\end{table}

\subsection{Baseline Model Performance Degradation Over Time}

\begin{figure}[!tbp]
  \centering
  \includegraphics[width=14cm]{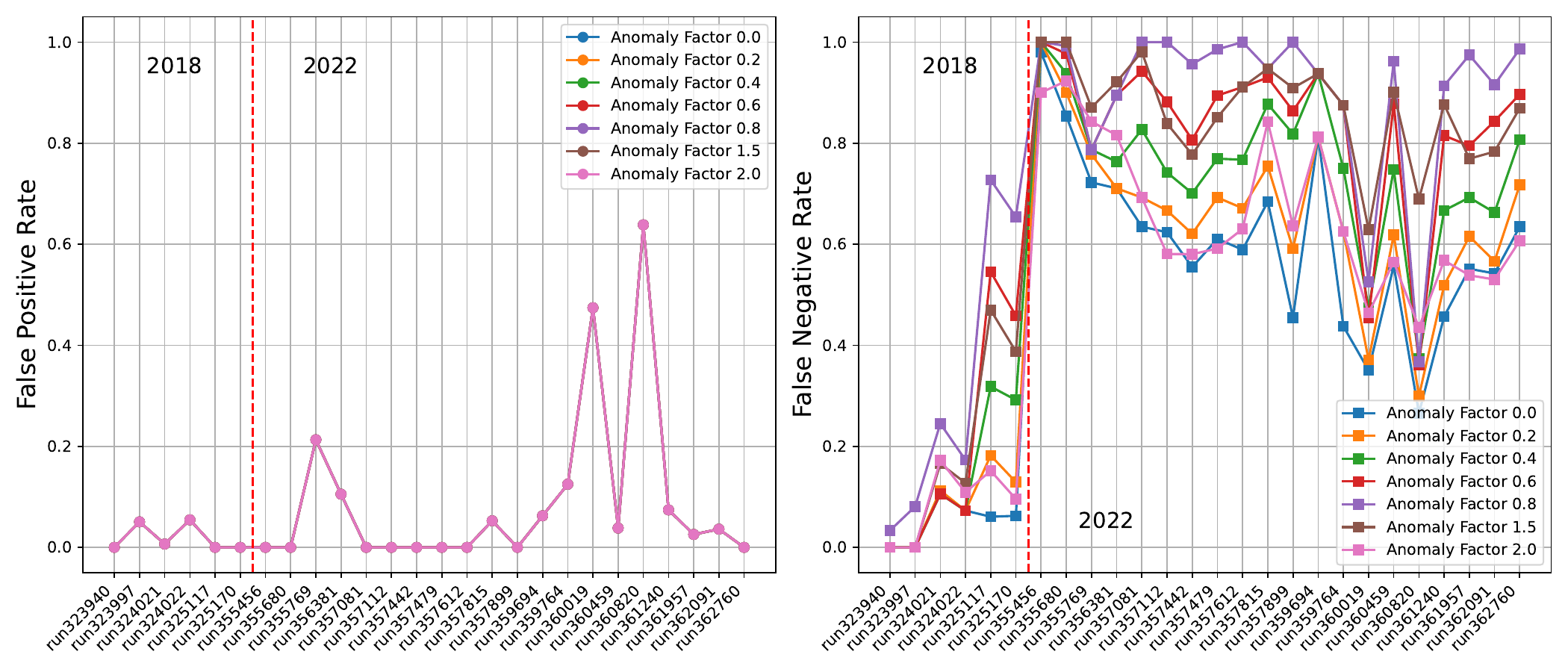}
  \caption{FPR (left) and FNR (right) of DepthViT model trained on \textit{Run323940} using the same hyperparameters detailed in Tab.~\ref{tab:anomaly_strength_metrics} when evaluated on test data from subsequent runs. Pointers indicate which dataset the model was evaluated on, while the red dashed line delineates runs originating in 2018 vs. 2022. Gradual model degradation is observed for 2018 data due to small shifts in the incoming data stream, while severe degradation is observed for 2022 data due to large shifts. \label{fig:Baseline_Model_Degradation}}
\end{figure}

In order to quantify the performance degradation that occurs due to shifts in the incoming data streams, the same baseline sub-ensemble model presented in Sec.~\ref{sec:baseline} was then evaluated on data originating from subsequent runs. Figure~\ref{fig:Baseline_Model_Degradation} presents the FPR and FNR of this model when evaluated on both its own validation data and subsequent 2018 and 2022 data for scenarios in which anomalies are present in each of the 5 LS considered within the time-window. It is observed that for subsequent runs originating in 2018 (\textit{Run323997} through \textit{Run325170}) which exhibit only small shifts in the incoming data streams, the model's FPR tends to decline while the FNR increases (a result which was referenced in Sec.~\ref{sec:methodology}). For runs originating in 2022 (\textit{Run355456} through \textit{Run362760}) which demonstrate large shifts in the incoming data streams, the model performs extremely poorly, failing to identify large numbers of anomalous events. The full performance metrics for this test are presented in Tab.~\ref{tab:Baseline_degradation}.

\begin{table}[ht]
\centering
\begin{tabular}{|c|c|c|c|c|c|}
\hline
\textbf{Anomaly Factor} & \textbf{Precision} & \textbf{Recall} & \textbf{F1 Score} & \textbf{FPR} & \textbf{FNR} \\
\hline
0.0 & 0.874 & 0.549 & 0.674 & 0.079 & 0.451 \\
0.2 & 0.863 & 0.498 & 0.631 & 0.079 & 0.503 \\
0.4 & 0.841 & 0.420 & 0.560 & 0.079 & 0.580 \\
0.6 & 0.810 & 0.338 & 0.477 & 0.079 & 0.662 \\
0.8 & 0.757 & 0.247 & 0.372 & 0.079 & 0.753 \\
1.5 & 0.800 & 0.316 & 0.453 & 0.079 & 0.684 \\
2.0 & 0.865 & 0.508 & 0.640 & 0.079 & 0.492 \\
\hline
\end{tabular}
\caption{Performance metrics for the baseline DepthViT model trained on \textit{Run323940} data, when deployed over all runs within the full dataset and implementing no continual learning mitigation techniques.}
\label{tab:Baseline_degradation}
\end{table}

\subsection{Model Ensembling \& Mitigation Techniques}

\begin{figure}[!tbp]
  \centering
  \includegraphics[width=14cm]{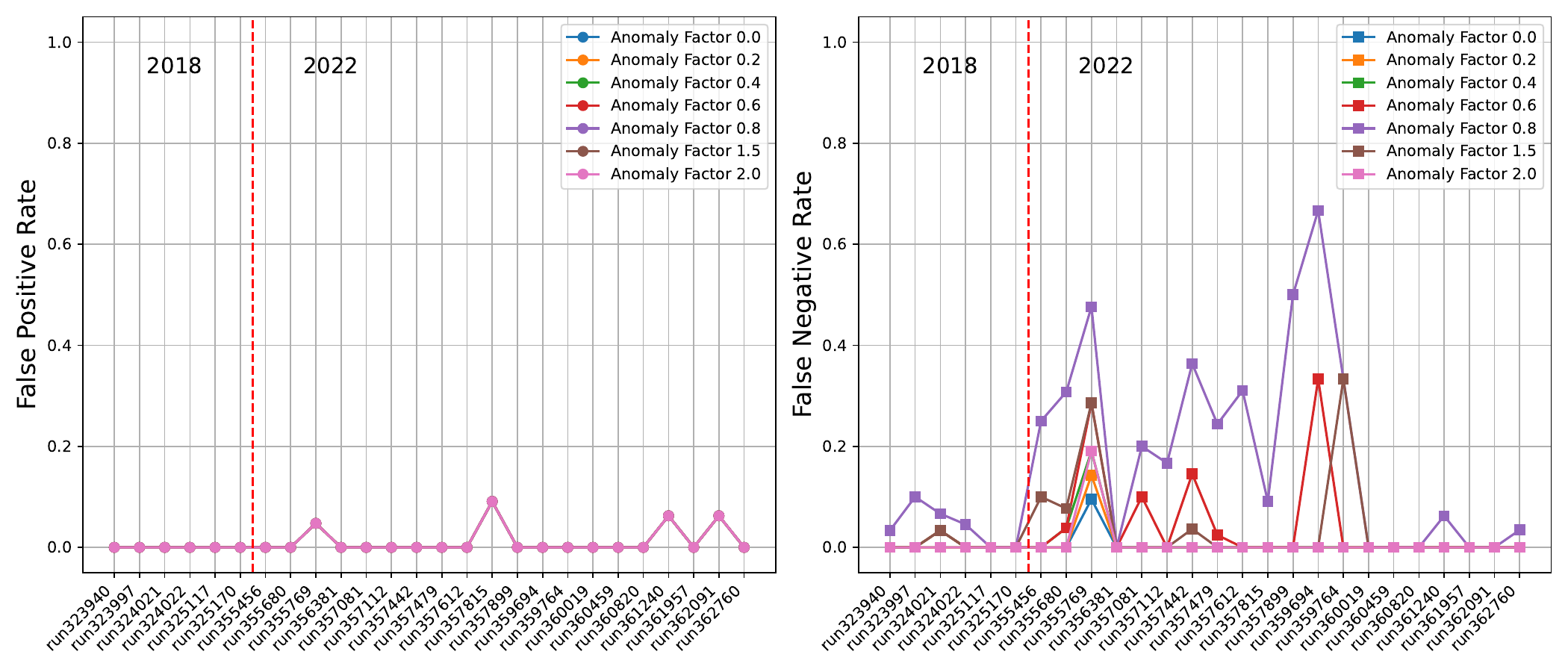}
  \caption{FPR (left) and FNR (right) of the baseline DepthViT model trained on \textit{Run323940}, using the same hyperparameters detailed in Tab.~\ref{tab:anomaly_strength_metrics} but with updated $\mu_{\mathrm{err}}$ and $\sigma_{\mathrm{err}}$ values derived from deployment on the validation data of the dataset being evaluated. Pointers indicate which test dataset the model was evaluated on, while the red dashed line delineates runs originating in 2018 vs. 2022. \label{fig:Baseline_Model_Stats_Update}}
\end{figure}

Multiple techniques for mitigating the observed model degradation were investigated and tested. This included periodically updating the $\mu_{\mathrm{err}}$ and $\sigma_{\mathrm{err}}$ values used in Eq.~\ref{eq:Z_score} via the latest data stream, ensembling together models trained on historic datasets with models trained on the most recent dataset, and the combination of these two techniques. Presented first are studies in which only $\mu_{\mathrm{err}}$ and $\sigma_{\mathrm{err}}$ values have been recalculated. For this test, the baseline sub-ensemble model from \textit{Run323940} was redeployed on test data from each subsequent run allowing for new $\mu_{\mathrm{err}}$ and $\sigma_{\mathrm{err}}$ statistics to be computed. These updated values were then applied during inference on the corresponding validation data which included synthetically injected anomalies. The model's weights were not updated and only the performance metrics used in the Z-score calculation were updated, implying that this does not constitute \textit{fine-tuning} of the model to the latest data. The resulting FPR and FNR values produced by this technique are shown in Fig.~\ref{fig:Baseline_Model_Stats_Update}. The result is that large shifts in the incoming data which would otherwise produce severe model degradation instead produce results characteristic with small shifts in the incoming data. The significance of this result is that because it does not require fine-tuning or further model training, it represents a potentially computationally resourceful mitigation technique that relies solely on the continued availability of pristine detector data. Table~\ref{tab:Stats_update_only} presents the relevant performance metrics from this study. It is observed that this techniques improves the FNR significantly, while improvements to the FPR are less substantial.

\begin{table}[ht]
\centering
\begin{tabular}{|c|c|c|c|c|c|}
\hline
\textbf{Anomaly Factor} & \textbf{Precision} & \textbf{Recall} & \textbf{F1 Score} & \textbf{FPR} & \textbf{FNR} \\
\hline
0.0 & 0.993 & 0.997 & 0.995 & 0.007 & 0.003 \\
0.2 & 0.993 & 0.995 & 0.994 & 0.007 & 0.005 \\
0.4 & 0.993 & 0.990 & 0.992 & 0.007 & 0.010 \\
0.6 & 0.993 & 0.968 & 0.981 & 0.007 & 0.032 \\
0.8 & 0.992 & 0.863 & 0.923 & 0.007 & 0.137 \\
1.5 & 0.993 & 0.977 & 0.985 & 0.007 & 0.024 \\
2.0 & 0.993 & 0.993 & 0.993 & 0.007 & 0.007 \\
\hline
\end{tabular}
\caption{Performance metrics for the baseline \textit{Run323940} model presented in Sec.~\ref{sec:baseline} when the relevant $\mu_{\mathrm{err}}$ and $\sigma_{\mathrm{err}}$ values have been updated using the validation data of the run being deployed on.}
\label{tab:Stats_update_only}
\end{table}

Presented next are results obtained by ensembling together multiple sub-ensemble models, while the $\mu_{\mathrm{err}}$ and $\sigma_{\mathrm{err}}$ values are left static and are not updated. For this study, no more than 4 models (each consisting of a quantile scaled and max scaled model in parallel) were ensembled together at a time.

The overall structure of the ensemble follows the flow chart detailed in Fig.~\ref{fig:Flow_chart}. The individual sub-ensemble models were never deployed on data that was used to train an individual model within the larger ensemble in order to avoid potential data contamination. Each time a new sub-ensemble model was introduced, the oldest sub-ensemble model was retired in order to maintain a fixed ensemble size. Table~\ref{tab:Ensemble_No_Update} shows the resulting performance metrics of this study. An important conclusion to note is that the static ensembling approach further improves the FNR values, however the FPR worsens compared to the baseline with no continual learning techniques employed (Tab.~\ref{tab:Baseline_degradation}).

\begin{table}[ht]
\centering
\begin{tabular}{|c|c|c|c|c|c|}
\hline
\textbf{Anomaly Factor} & \textbf{Precision} & \textbf{Recall} & \textbf{F1 Score} & \textbf{FPR} & \textbf{FNR} \\
\hline
0.0 & 0.779 & 1.000 & 0.876 & 0.283 & 0.000 \\
0.2 & 0.779 & 1.000 & 0.876 & 0.283 & 0.000 \\
0.4 & 0.779 & 0.998 & 0.875 & 0.283 & 0.002 \\
0.6 & 0.777 & 0.983 & 0.868 & 0.283 & 0.017 \\
0.8 & 0.762 & 0.908 & 0.829 & 0.283 & 0.092 \\
1.5 & 0.777 & 0.988 & 0.870 & 0.283 & 0.012 \\
2.0 & 0.779 & 1.000 & 0.876 & 0.283 & 0.000 \\
\hline
\end{tabular}
\caption{Performance metrics for an ensemble of models generated on 4 total runs, where a new set of models were introduced every run of the dataset. The older models within the ensemble do not have their $\mu_{\mathrm{err}}$ and $\sigma_{\mathrm{err}}$ values updated and instead remain static.}
\label{tab:Ensemble_No_Update}
\end{table}

Finally, a study was performed which combined both of these techniques, in which a new sub-ensemble model was trained and added to the ensemble for every run, while the $\mu_{\mathrm{err}}$ and $\sigma_{\mathrm{err}}$ values were updated for all individual sub-ensemble models within the ensemble using the validation data of the latest run. The entire ensemble was then evaluated using the latest run's test data. The result, shown in Tab.~\ref{tab:Update_every_run}, is an ensemble which exhibits $>98\%$ precision across all anomaly factors tested, and similar recall performance for all anomaly factors except the most subtle tested (0.8). Most importantly, both the FPR and FNR values are significantly improved relative to the application of either technique alone, and relative to baseline performance which included no continual learning mitigation techniques.

\begin{table}[ht]
\centering
\begin{tabular}{|c|c|c|c|c|c|}
\hline
\textbf{Anomaly Factor} & \textbf{Precision} & \textbf{Recall} & \textbf{F1 Score} & \textbf{FPR} & \textbf{FNR} \\
\hline
0.0 & 0.988 & 1.000 & 0.994 & 0.012 & 0.000 \\
0.2 & 0.988 & 0.998 & 0.993 & 0.012 & 0.002 \\
0.4 & 0.988 & 0.995 & 0.992 & 0.012 & 0.005 \\
0.6 & 0.988 & 0.985 & 0.987 & 0.012 & 0.015 \\
0.8 & 0.987 & 0.891 & 0.937 & 0.012 & 0.109 \\
1.5 & 0.988 & 0.990 & 0.989 & 0.012 & 0.010 \\
2.0 & 0.988 & 0.997 & 0.993 & 0.012 & 0.003 \\
\hline
\end{tabular}
\caption{Performance metrics for an ensemble of models generated on 4 total runs, where a new set of models were introduced for every run of the dataset.}
\label{tab:Update_every_run}
\end{table}

\subsection{Model Ensembling Performance Compared to Latest Model Performance}

A final study was performed to compare the performance of an ensembled set of models to that of a single model only trained on the latest data stream. To test this, ensembling of older models was employed, but instead a model was generated and employed for every run, and then immediately discarded. This would represent a scenario in which no older models were able to contribute to the performance of the overall ensemble. These results are presented in Tab.~\ref{tab:Single_model_every_run}. It is observed that ensembling provides an improvement of 100\% and 55\%  to the FNR for anomaly factors of 0.0 and 1.5, respectively (strong anomalies), and an improvement of 11\% to the FNR for anomaly factor 0.8 (subtle anomalies). These results indicate that the ensembling improves the overall performance and stability of the system.

\begin{table}[ht]
\centering
\begin{tabular}{|c|c|c|c|c|c|}
\hline
\textbf{Anomaly Factor} & \textbf{Precision} & \textbf{Recall} & \textbf{F1 Score} & \textbf{FPR} & \textbf{FNR} \\
\hline
0.0 & 0.997 & 0.998 & 0.998 & 0.003 & 0.002 \\
0.2 & 0.997 & 0.993 & 0.995 & 0.003 & 0.007 \\
0.4 & 0.997 & 0.988 & 0.992 & 0.003 & 0.012 \\
0.6 & 0.997 & 0.972 & 0.984 & 0.003 & 0.029 \\
0.8 & 0.996 & 0.878 & 0.933 & 0.003 & 0.122 \\
1.5 & 0.997 & 0.978 & 0.987 & 0.003 & 0.022 \\
2.0 & 0.997 & 0.997 & 0.997 & 0.003 & 0.003 \\
\hline
\end{tabular}
\caption{Performance metrics for an ensemble of models consisting of 2 total models (one for each employed scaling technique) trained only on the latest run.}
\label{tab:Single_model_every_run}
\end{table}

\section{CONCLUSION}

In this work, DepthViT, a novel, lightweight ML architecture, has been introduced and evaluated on the task of anomaly detection in HEP data. The DepthViT architecture varies from traditional vision transformers in that input channels do not share kernel filters, and that the attention mechanism is applied depthwise rather than along the input sequence. Two continual learning approaches have been evaluated and applied. The first was an updating of the evaluation statistics used to perform a Z-score calculation using data drawn from the latest input stream, while the second was a straightforward ensembling scheme which ensembled together models trained on prior data, in addition to the latest input data stream. It was observed that employing both of these techniques together improved the performance of the overall system when compared to individual models and also made the system more resilient to changes in the incoming data stream. These findings suggest that lightweight, modular ensemble architectures can provide a practical approach to reducing the effects of model degradation without the need for retraining on historical data. The broader implications of this work extends beyond HEP, as the same continual adaptation strategy can be directly applied to industrial monitoring environments where data naturally evolves over time, such as manufacturing lines with aging sensors, or the emergence of new failure modes in advanced manufacturing processes. Additionally, the novel DepthViT architecture introduced could be extended to applications in other branches of research where critical information for the task is patch-local but where channels do not necessarily correspond to the same physical point in space. Example tasks could include classification and regression based on key peaks in spectral data, or local feature extraction from multi-channel optical data.

\section{Code Availability}
The code for this work is made available at the following link along with simulated demo data for testing and development: \url{https://github.com/daleadamjulson/DepthViT_CML}.

\begin{acknowledgments}
This work was supported in part by a Phase I contract from the U.S. Department of Energy Small Business Technology Transfer Program (contract number DE-SC0024776). Individuals have received support from the Belgian Fonds de la Recherche Scientifique, and Fonds voor Wetenschappelijk Onderzoek; the Brazilian Funding agencies (CNPq, CAPES, FAPERJ, FAPERGS, and FAPESP); SRNSF (Georgia); the Bundesministerium f\"ur Bildung und Forschung, the Deutsche Forschungsgemeinschaft (DFG), under Germany's Excellence Strategy -- EXC 2121 ``Quantum Universe'' -- 390833306, and under project number 400140256 - GRK2497, and Helmholtz-Gemeinschaft Deutscher Forschungszentren, Germany; the National Research, Development and Innovation Office (NKFIH) (Hungary) under project numbers K~128713, K~143460, and TKP2021-NKTA-64; the Department of Atomic Energy and the
Department of Science and Technology, India; the Ministry of Science, ICT and Future Planning, and National Research Foundation (NRF), Republic of Korea; the Lithuanian Academy of Sciences; the Scientific and Technical Research Council of Turkey, and Turkish Energy, Nuclear and Mineral Research Agency; the National Academy of Sciences of Ukraine; the US Department of Energy.
\end{acknowledgments}

\printbibliography[title={REFERENCES }, heading=bibintoc]  % Gunakan judul LITERATURE

\clearpage
\section*{The CMS HCAL Collaboration}
\addcontentsline{toc}{section}{The CMS HCAL Collaboration}
\input{authorlist.tex}

\end{document}

%% file: authorlist.tex
\parindent 0pt
\parskip 4pt
\newcommand{\cmsinstitute}[1]{\par\pagebreak[3]\bfseries #1 \mdseries\\[0pt]}
\newcommand{\cmsorcid}[1]{\href{https://orcid.org/#1}{\hspace*{0.1em}\raisebox{-0.45ex}{\includegraphics[width=1em]{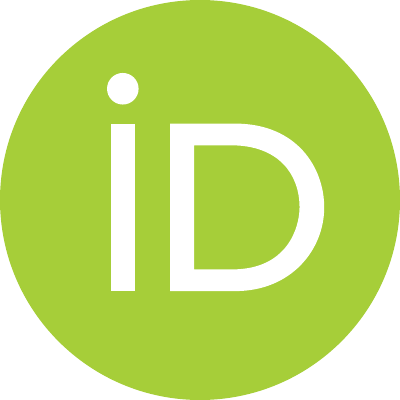}}}}
\makeatletter
\newcommand{\cmsAuthorMark}[1]{\hbox{\@textsuperscript{\normalfont#1}}}
\makeatother
\newskip{\cmsinstskip} \cmsinstskip=4pt

%\cmsinstitute{The CMS HCAL Collaboration}

\cmsinstitute{Yerevan~Physics~Institute, Yerevan, Armenia}
{\tolerance=6000
A.~Gevorgyan, A.~Petrosyan, A.~Tumasyan
\par}

\cmsinstitute{Universiteit~Antwerpen, Antwerpen, Belgium\cmsAuthorMark{*}}
{\tolerance=6000
M.~Van~De~Klundert, H.~Van~Haevermaet, P.~Van~Mechelen\cmsorcid{0000-0002-8731-9051}, A.~Van~Spilbeeck
\par}

\cmsinstitute{Centro~Brasileiro~de~Pesquisas~Fisicas, Rio de Janeiro, Brazil\cmsAuthorMark{*}}
{\tolerance=6000
G.A.~Alves\cmsorcid{0000-0002-8369-1446}, C.~Hensel\cmsorcid{0000-0001-8874-7624}
\par}

\cmsinstitute{Universidade~do~Estado~do~Rio~de~Janeiro, Rio de Janeiro, Brazil\cmsAuthorMark{*}}
{\tolerance=6000
W.L.~Ald\'{a}~J\'{u}nior\cmsorcid{0000-0001-5855-9817}, W.~Carvalho\cmsorcid{0000-0003-0738-6615}, J.~Chinellato\cmsAuthorMark{1}, C.~De~Olivera~Martins, D.~Matos~Figueiredo, C.~Mora~Herrera\cmsorcid{0000-0003-3915-3170}, H.~Nogima\cmsorcid{0000-0001-7705-1066}, W.L.~Prado~Da~Silva, E.J.~Tonelli~Manganote, A.~Vilela~Pereira\cmsorcid{0000-0003-3177-4626}
% ...
\par}

\vskip\cmsinstskip

\dag:~Deceased\\
$^{*}$No longer in CMS HCAL Collaboration\\
$^{1}$Also at Universidade Estadual de Campinas, Campinas, Brazil\\
% ...

%\end{document}